\documentclass[aip,reprint]{revtex4-1}
\usepackage{graphicx}
\usepackage{ulem}
\usepackage{amssymb}
\usepackage{amsmath}
\usepackage{wasysym}
\usepackage{relsize}
\begin{document}


\newcommand{\be}{\begin{equation}}
\newcommand{\ee}{\end{equation}}
\newcommand{\bea}{\begin{eqnarray}}
\newcommand{\eea}{\end{eqnarray}}
\newcommand{\Tbar}{{\bar{T}}}
\newcommand{\En}{{\cal E}}
\newcommand{\K}{{\cal K}}
\newcommand{\U}{{\cal U}}
\newcommand{\GC}{{\cal \tt G}}
\newcommand{\Lop}{{\cal L}}
\newcommand{\DB}[1]{\marginpar{\footnotesize DB: #1}}
\newcommand{\q}{\vec{q}}
\newcommand{\kt}{\tilde{k}}
\newcommand{\Lopn}{\tilde{\Lop}}
\newcommand{\noi}{\noindent}
\newcommand{\ovn}{\bar{n}}
\newcommand{\ovx}{\bar{x}}
\newcommand{\ovE}{\bar{E}}
\newcommand{\ovV}{\bar{V}}
\newcommand{\ovU}{\bar{U}}
\newcommand{\ovJ}{\bar{J}}
\newcommand{\calE}{{\cal E}}
\newcommand{\ovphi}{\bar{\phi}}
\newcommand{\zt}{\tilde{z}}
\newcommand{\ttl}{\tilde{\theta}}
\newcommand{\nuv}{\rm v}
\newcommand{\ds}{\Delta s}
\newcommand{\fn}{{\small {\rm  FN}}}
\newcommand{\cc}{{\cal C}}
\newcommand{\cd}{{\cal D}}
\newcommand{\tth}{\tilde{\theta}}
\newcommand{\cb}{{\cal B}}
\newcommand{\cg}{{\cal G}}
\newcommand\norm[1]{\left\lVert#1\right\rVert}

\title{The Schottky Conjecture and beyond}


\author{Debabrata Biswas}
\affiliation{
Bhabha Atomic Research Centre,
Mumbai 400 085, INDIA}
\affiliation{Homi Bhabha National Institute, Mumbai 400 094, INDIA}


\begin{abstract}
  The `Schottky Conjecture' deals with the electrostatic field enhancement at the tip of compound
  structures such as a hemiellipsoid on top of a hemisphere. For such a  2-primitive  compound structure,
  the apex field enhancement factor  $\gamma_a^{(C)}$ is conjectured to be multiplicative
  ($\gamma_a^{(C)} = \gamma_a^{(1)} \gamma_a^{(2)}$)  provided the structure
  at the base (labelled 1, e.g. the hemisphere) is much larger than the structure on top
  (referred to as crown and labelled 2, e.g. the hemi-ellipsoid).
  We first demonstrate numerically that for generic smooth structures,
  the conjecture holds
  in the limiting sense when the apex radius of curvature of the primitive-base $R_a^{(1)}$,
  is much larger than the height of the crown $h_2$ (i.e. $h_2/R_a^{(1)} \rightarrow 0$).
  If the condition is somewhat relaxed, we show that it is the electric field
  above the primitive-base (i.e. in the absence of the crown),
  averaged over the height of the crown,
  that gets magnified instead of the field at the apex of the primitive-base.
  This observation leads to the Corrected
  Schottky Conjecture (CSC), which for 2-primitive structures reads as
  $\gamma_a^{(C)}\simeq \langle \gamma_a^{(1)}\rangle\gamma_a^{(2)}$
  where $\langle  . \rangle$ denotes the average value over the height of the crown.
  For small protrusions ($h_2/h_1$ typically less than 0.2), $\langle \gamma_a^{(1)}\rangle$ can
  be approximately determined using the Line Charge Model so that
  $\gamma_a^{(C)} \simeq \gamma_a^{(1)} \gamma_a^{(2)} (2R_a^{(1)}/h_2)\ln(1 + h_2/2R_a^{(1)})$. The error
  is found to be within $1\%$ for $h_2/R_a^{(1)} < 0.05$, increasing to about $3\%$ (or less) for $h_2/R_a^{(1)} = 0.1$
  and bounded below 5\% for $h_2/R_a^{(1)}$ as large as  0.5. The CSC is
  also found to give good results for 3-primitive compound structures.
  The relevance of the Corrected Schottky Conjecture for field emission is discussed.
\end{abstract}

\maketitle

\section{Introduction}
\label{sec:intro}

In 1923, Schottky  argued that the apex field enhancement factor (AFEF) at
the tip of a compound structure should be a product of the AFEF values of the
successive primitive structures comprising it,
provided each structure is much smaller than the preceding one \cite{Schott23,stern}.
Nearly a hundred years since, there
is renewed interest \cite{miller07,miller09,jensen16,deAssis16,marcelino,harris19a,harris19b,zhu2019} in this
conjecture for various reasons. A primary
cause of breakdown in vacuum devices is thought to be electron emission from micro-protrusions on
an otherwise smooth surface on application of a DC or RF field.
Since, appreciable electron emission requires electric field strengths upwards of
$3$GV/m, it is now accepted that micro-protrusions can have very high field enhancement factors due
to the compounding effect suggested by Schottky \cite{Schott23}. The conjecture is also relevant in
situations where field emission is desirable since, even though we now have a fair idea about the
field enhancement factor of generic single primitive structures \cite{forbes2003,db18_fef,db_sg,db_anode}, a simple and
useful model of compound structures is yet to be formulated.
An analytical formula (even an empirical one) for the AFEF of compound structures would
no doubt be extremely useful in optimizing the field emission current with respect to the parameters of
single emitters and might well be of future use in studying large area  field emitters \cite{db_rudra18,db_rudra19}
of compound entities. It is thus necessary that we revisit the Schottky Conjecture (SC)
using the tools presently at our disposal
and try to go beyond in situations that do not quite obey the stringent requirements
of the conjecture.

The apex field enhancement factor $\gamma_a$ is defined as the ratio of the
local field at the apex ($E_a$) and the
macroscopic field far away from the emitter ($E_0$). In a planar diode configuration with the cathode
plate at $z = 0$ and the anode plate at $z = D$ having a potential difference $V$ with respect to
the cathode plate, the macroscopic field $E_0 = V/D$. Consider now an axially symmetric
curved emitter of height $h$ (with $h << D$) and apex radius of curvature $R_a$ placed normal to the
cathode plate. The local field at the apex $E_a = \gamma_a E_0$ where $\gamma_a$ is the apex
field enhancement factor of an isolated emitter with the anode far away \cite{anode_position}.

Consider now two structures having AFEF $\gamma_a^{(1)}$ and $\gamma_a^{(2)}$ respectively. When the second
structure is mounted on top of the first, the compound structure does not necessarily have its AFEF
value as $\gamma_a^{(1)} \gamma_a^{(2)}$. However, if the apex radius of curvature of the first structure
$R_a^{(1)}$ is much larger compared to the height $h_2$ of the second structure, the AFEF
of the compound structure can be closely approximated by $\gamma_a^{(1)} \gamma_a^{(2)}$ since
the second structure finds itself on a quasi-planar base  and takes advantage of the
enhanced local field near the apex of the first structure.
Thus, at a basic level, the Schottky Conjecture seems plausible provided $h_2/R_a^{(1)} \rightarrow 0$.
If the system has more than 2 primitive structures, a similar logic would imply that the
AFEF of the compound structure be a product of the primitive AFEF values provided $h_{i+1}/R_a^{(i)} \rightarrow 0$
for all $i$. Thus the AFEF value of the compound structure having $N$ primitives may be expressed as
$\gamma_a^{(C)} = \prod_{i=1}^N \gamma_a^{(i)}$.

\begin{figure}[bth]
  \begin{center}
\vskip -0.75cm
\hspace*{0.250cm}\includegraphics[width=0.43\textwidth]{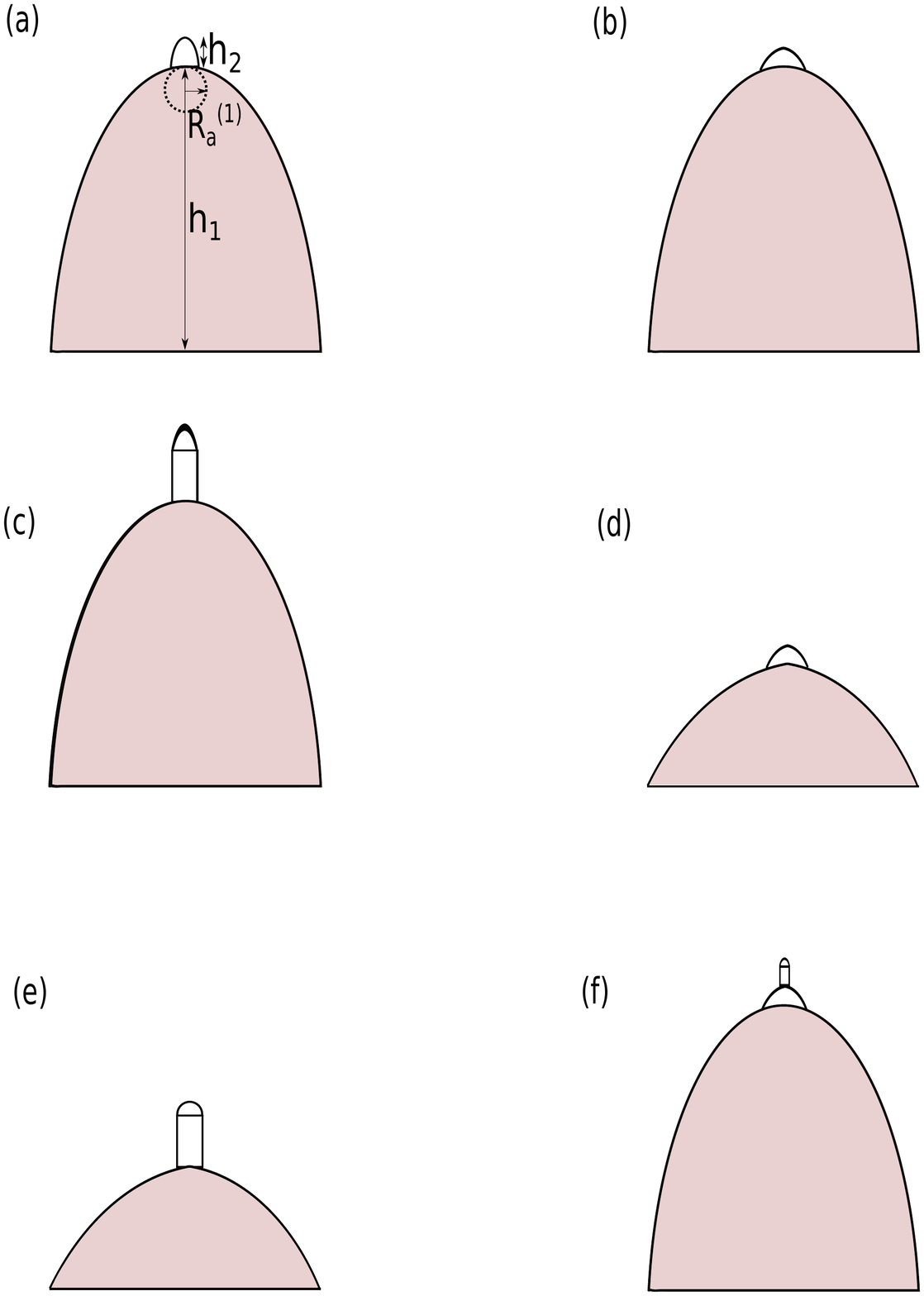}
\vskip -0.25cm
\caption{A 2-dimensional schematic of 6 axially symmetric compound structures 
  considered here. The five 2-primitive compound structures are
  (a) an ellipsoid on ellipsoid (b) a paraboloid on ellipsoid
  (c) a hemiellipsoid-on-cylindrical-post (HECP) on ellipsoid (d) a paraboloid on paraboloid
  (e) a hemisphere-on-cylindrical post (HCP) on paraboloid, while (f) is a 3-primitive structure
  consisting of an HECP on a paraboloid which is mounted on an ellipsoid.  The top-most
  primitive structure in each case is referred to as the crown. }
\vskip -.75cm
\label{fig:para_on_ellip}
\end{center}
\end{figure}

In practice, compound structures may have $h_{i+1}/R_a^{(i)}$ small but non-zero and it would
be desirable in such cases to have a Corrected Schottky Conjecture with correction terms \cite{see_miller}
expressed in terms of $h_{i+1}/R_a^{(i)}$. 
In other situations, $h_{i+1}/R_a^{(i)}$ may
be much larger than unity and even though the Schottky Conjecture is clearly inapplicable,
there is need to model such compound structures in terms of the primitive structures, at least
in an approximate way. In either case, it is necessary to go beyond the Schottky conjecture
in order to determine the AFEF of a compound structure in terms of the primitive AFEF values with some
degree of accuracy. The paper is devoted to this endeavour and while no proof is provided, plausibility
arguments for a Corrected Schottky Conjecture (CSC) is given along with numerical verification
that help us in establishing
the usefulness of the CSC for practical compound structures.

Section \ref{sec:corrected_schott} deals with the Corrected Schottky Conjecture that may be
applied to compound structures where $h_{i+1}/R_a^{(i)}$ is not vanishingly small. This is
demonstrated numerically in section \ref{sec:numerical}  for several compound structures and
a discussion on its relevance to field emission is provided in section \ref{sec:discussions}.
Unless otherwise stated, all numerical calculations presented here have
been performed using the AC/DC module of COMSOL Multiphysics software v5.4.

\section{The Corrected Schottky Conjecture}
\label{sec:corrected_schott}

In order to appreciate the need for correction terms to the Schottky Conjecture, we shall
consider compound shapes made up of two primitive structures such as hemiellipsoids, paraboloids
or a hemiellipsoid on a cylindrical post (HECP). Fig.~\ref{fig:para_on_ellip} shows 6 compound
structures of which the first 5 [(a) to (e)] are 2-primitive structures.

\begin{figure}[bth]
  \begin{center}
\vskip -0.5cm
\hspace*{-0.750cm}\includegraphics[width=0.575\textwidth]{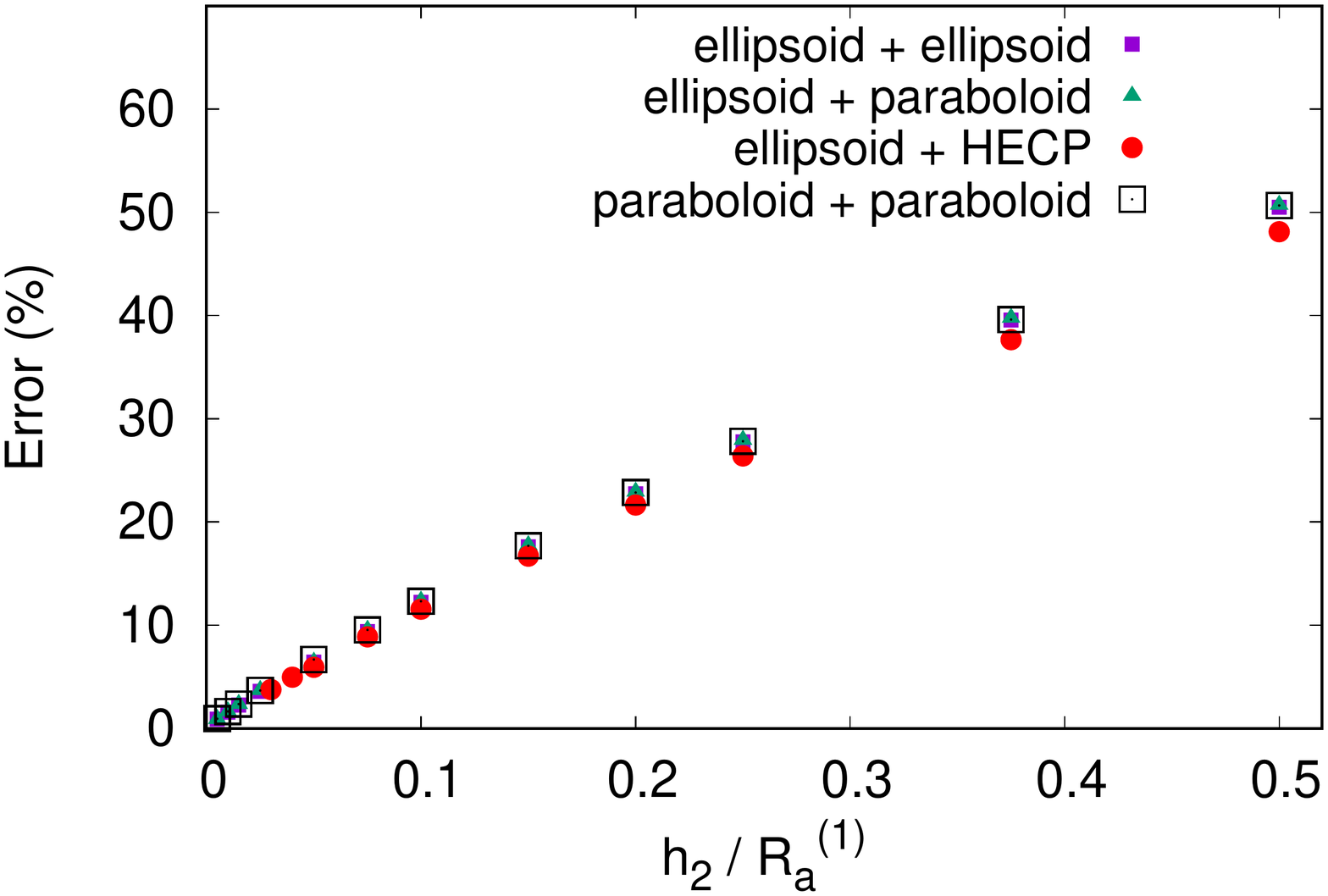}
\vskip -0.75cm
\caption{The error, as a defined in Eq.~(\ref{eq:err1}), as a function of $h_2/R_a^{(1)}$ for 4 compound shapes,
  each having 2 primitives as shown in (a)-(d) of Fig.~\ref{fig:para_on_ellip}. The error in the
  Schottky Conjecture grows as $h_2/R_a^{(1)}$ increases, and is roughly the same for all shapes.}
\vskip -0.5cm
\label{fig:error_schott}
\end{center}
\end{figure}

Fig.~\ref{fig:error_schott} shows the relative error in AFEF
value of the compound structure (denoted by $\gamma_a^{(C)}$) defined as

\be
\text{Error} = \frac{\gamma_a^{(C)} - \gamma_a^{(1)} \gamma_a^{(2)}}{\gamma_a^C} \times 100  \label{eq:err1}
\ee

\noi
for 4 different combinations of primitives [(a) to (d) of Fig.~\ref{fig:para_on_ellip}].
In each case the apex radius of curvature of the base, $R_a^{(1)} = 1\mu$m while the apex radius of the crown
is $R_a^{(2)} = 5$nm.
The first is a hemiellipsoid on top of another
(larger) hemiellipsoid, the second is a paraboloid on top of an hemiellipsoid, the third an HECP structure
on top of a hemiellipsoid and finally, we consider a paraboloid on top of another paraboloid. In the
first 3 cases, the total height of the compound structure is $3\mu$m while the total height of the
last case is $6\mu$m. The quantities $\gamma_a^{(C)}, \gamma_a^{(1)}$ and $\gamma_a^{(2)}$ have been evaluated using
COMSOL.
In all cases, the error grows with $h_2/R_a^{(1)}$. It is around 12\% when $h_2/R_a^{(1)} = 0.1$, and is as high as
50\% when $h_2/R_a^{(1)} = 0.5$. Thus, the Schottky Conjecture seems to holds good when
$h_2/R_a^{(1)}$ is very small but the error grows as the protrusion on top of the first structure
increases in height compared to the apex radius of curvature $R_a^{(1)}$.
Note that despite the variation in the composition of the compound structures, the error is
roughly the same for all shapes. It is thus obvious that a correction to the Schottky Conjecture must
principally be a function of $h_2/R_a^{(1)}$.

Clearly, the compounding process wherein a smaller structure (crown) sits on top of a larger base, leads to an
amplification of the local field. The equipotential curves that existed close to the apex of the
primitive-base (in the
absence of the crown), now have to cling to the crown and get further compressed due to the
enhancing effect of the smaller structure. When the height of the crown ($h_2$) is vanishingly small,
the local field that existed at the apex of the
primitive-base ($E_0 \gamma_a^{(1)}$) gets amplified to $E_0 \gamma_a^{(1)} \gamma_a^{(2)}$ at the apex of the crown.
However, as the height of the crown gets larger, it is not obvious whether the field that get
amplified by the factor $\gamma_a^{(2)}$ is the one at the apex of the primitive-base or one that
depends on the height of the crown.

To test this, consider the 2-primitive compound structure consisting of a hemiellipsoid as the base
($R_a^{(1)} = 1000$nm, $h_1 = 2900$nm, $\gamma_a^{(1)} = 4.84$)  and a hemiellipsoid on a cylindrical post (HECP)
as the crown ($R_a^{(2)} = 10$nm). The height of the HECP ($h_2$) is varied from 200nm to 7100nm.
Thus $\gamma_a^{(2)}$ varies as does the AFEF of the compound structure.
We are interested in the quantity $\gamma_a^{(C)}/\gamma_a^{(2)}$ where $\gamma_a^{(C)}$ corresponds
to the apex field enhancement factor of the compound structure while $\gamma_a^{(2)}$ is
the AFEF of the HECP (i.e. the crown). In the limit $h_2/R_a^{(1)} \rightarrow 0$,
$\gamma_a^{(C)}/\gamma_a^{(2)}$ should approach the AFEF of the base (i.e. $\gamma_a^{(1)} = 4.84$).
For all other values it should give an effective enhancement factor corresponding to the
field that get amplified by the crown.

\begin{figure}[bth]
  \begin{center}
\vskip -0.75cm
\hspace*{-0.750cm}\includegraphics[width=0.575\textwidth]{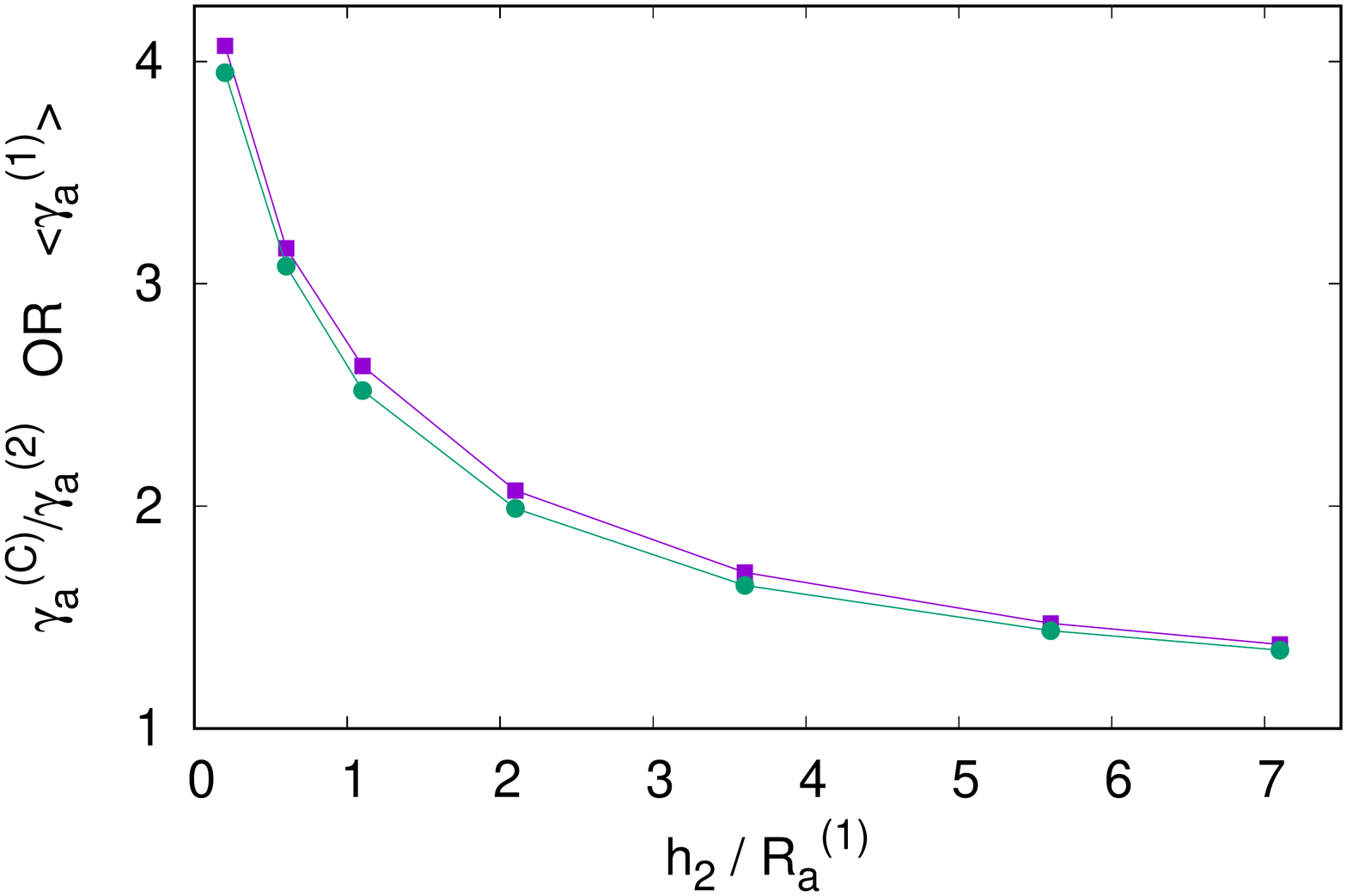}
\vskip -0.5cm
\caption{The values of $\gamma_a^{(C)}/\gamma_a^{(2)}$ (solid circles) is plotted alongside
 $\langle \gamma_a^{(1)} \rangle$ (solid squares) defined by Eq.~(\ref{eq:av_gam}).}
\vskip -0.5cm
\label{fig:average_field}
\end{center}
\end{figure}

Fig.~\ref{fig:average_field} shows the values of $\gamma_a^{(C)}/\gamma_a^{(2)}$
at a few values of $h_2/R_a^{(1)}$ (denoted by circles). As mentioned, in the limit $h_2 \rightarrow 0$,
$\gamma_a^{(C)}/\gamma_a^{(2)} \rightarrow \gamma_a^{(1)}$. Indeed Fig.~\ref{fig:average_field} shows such
a trend while as $h_2$ becomes large, $\gamma_a^{(C)}/\gamma_a^{(2)} \rightarrow 1$. Thus, the field
that gets amplified by the factor $\gamma_a^{(2)}$ is smaller than the field at the
apex of the primitive-base for non-zero values of $h_2$.
Note that the field at a height $h_2$ above the apex of the primitive
base is close to $E_0$ when $h_2 \gtrapprox 2 R_a^{(1)}$. Thus, $E_0 \gamma_a^{(C)}/\gamma_a^{(2)}$
is smaller than the field at the apex of the primitive-base but larger than the
field at a height $h_2$ above the apex of the primitive-base. As seen in Fig.~\ref{fig:average_field},
$1 \leq \gamma_a^{(C)}/\gamma_a^{(2)} \leq \gamma_a^{(1)}$.

Fig.~\ref{fig:average_field} also shows the averaged quantity (denoted by squares)

\be
\langle \gamma_a^{(1)} \rangle = \frac{ \frac{1}{h_2} \int_{h_1}^{h_1+ h_2} E_z(z) dz}{E_0} = \frac{\langle E_z(z)\rangle}{E_0} \label{eq:av_gam}
\ee

\noi
where $h_1$ is the height of the  primitive-base and $E_z(z)$ the field along the axis,
calculated here using COMSOL.
Clearly $\langle \gamma_a^{(1)} \rangle$ follows $\gamma_a^{(C)}/\gamma_a^{(2)}$ closely over a range of
$h_2/R_a^{(1)}$. Thus, it is the average field above the base that gets amplified by the crown
by a factor $\gamma_a^{(2)}$. This leads us to the Corrected Schottky Conjecture (CSC):

\be
\gamma_a^{(C)} \simeq \langle \gamma_a^{(1)}\rangle \gamma_a^{(2)}
\ee

\noi
for a 2-primitive system, while for an $N$ primitive compound structure

\be
\gamma_a^{(C)} \simeq  \gamma_a^{(N)} \Pi_{i=1}^{N-1} \langle \gamma_a^{(i)}\rangle  \label{eq:gencsc}
\ee

\noi
where the averaging at the $i^{\text{th}}$ primitive stage is over the height $h_{i+1}$ of the next stage.
Eq.~(\ref{eq:gencsc}) is a useful approximation when successive stages are not limited by the
smallness criterion. Rather $h_{i+1}/R_a^{(i)}$ may in fact be larger than 1 as illustrated in
Fig.~\ref{fig:average_field}. The CSC  as in Eq.~(\ref{eq:gencsc}) has been similarly tested for other
compound structures. As another example, the base is considered to be a paraboloid of
radius of curvature $R_a^{(1)} = 1\mu$m and height $h_1 = 5.525\mu$m while the crown is a hemisphere on
a cylindrical post (HCP) with $R_a^{(2)} = 5$nm and total height $h_2 = 475$nm (see schematic (e) of
Fig.~\ref{fig:para_on_ellip}). For this system,
the average error in CSC prediction is about $2\%$ while the error in Schottky Conjecture is about
$45\%$.

Apart from special primitive-bases such as the hemi-ellipsoid, the exact axial electrostatic field is in
general unknown and hence $\langle \gamma_a^{(i)}\rangle$ can only be evaluated numerically, for
instance using COMSOL.
However, when $h_{i+1}$ is smaller than $h_i$, it is possible to express the electrostatic
field approximately using the nonlinear line charge model \cite{jap16,cosine,db_anode,mesa,pogo2009,harris15}
so that a simple useful approximation for $\gamma_a^{(C)}$ can be arrived at.

The field $E_z$, above the apex of a generic axially symmetric emitter of apex radius of
curvature $R_a^{(1)}$ and height $h_1$, can be approximately expressed using the nonlinear
line charge model as (see Eq.~(38) in [\onlinecite{cosine}] with $\rho = 0$)

\be
E_z \simeq \frac{f(L)}{4\pi\epsilon_0} \frac{2zL}{z^2 - L^2} (1 - {\cal C})  \label{eq:Ez0}
\ee

\noi
where $z$ is measured from the cathode plane, $L$ represents the extent of the line charge with
$L^2 \simeq h_1(h_1 - R_a^{(1)})$, $f(L)$ is related to the line charge density and
${\cal C}$ is a correction term that is unknown {\it a priori} and arises from the
nonlinearity in line charge distribution. It (i.e. ${\cal C}$) is zero
for a hemiellipsoid (linear line charge) and in general varies with the shape of the emitter.
Importantly, for sharp emitters ($h_1/R_a^{(1)} >> 1$) the correction is in general small.
Even though the primitive-bases considered here are not necessarily sharp,
we shall neglect ${\cal C}$ hereafter as a reasonable approximation.
Note also that $2zL/(z^2 - L^2)$ in Eq.~(\ref{eq:Ez0}) is the leading term and there exists a logarithmic
correction\cite{cosine} ($\propto \ln((z+L)/(z-L))$) which cannot be neglected for larger $z$.
For sharp emitters where $L \simeq h_1$, the logarithmic contribution is less than half for $z = 1.2h_1$
so that Eq.~(\ref{eq:Ez0}) can be considered to be reasonably valid for $z < 1.2h_1$.

Writing $z = h_1 + \Delta$, and using\cite{db18_fef,cosine}

\be
\frac{f(L)}{4\pi\epsilon_0} \frac{2h_1L}{h_1^2 - L^2} \simeq \gamma_a^{(1)} E_0
\ee

\noi
the above equation can be expressed as 

\be
E_z(\Delta) \simeq E_0 \gamma_a^{(1)} (1 - {\cal C}) \frac{h_1 + \Delta}{h_1}\frac{1}
{1 + \frac{2 h_1 \Delta + \Delta^2}{h_1 R_a^{(1)}}}.
\ee

\noi
For small protrusions ($h_2/h_1 < 0.2$) from the primitive-base, $\Delta/R_a^{(1)}$
as well as $\Delta/h_1$ are small. Thus, the field along the axis for $\Delta < h_1$ is

\be
E_z(\Delta) \simeq E_0 \gamma_a^{(1)}  \frac{1}{1 + \frac{2\Delta}{R_a^{(1)}}}. \label{eq:Ezfinal}
\ee

\noi
Eq.~(\ref{eq:Ezfinal}) serves as a reasonable approximation
for calculating $\langle \gamma_a^{(1)} \rangle$ even though, the neglect of ${\cal{C}}$ and the logarithmic term are
sources of minor errors.

The electric field averaged over the height of the crown is

\be
\langle E_z \rangle  \simeq E_0 \gamma_a^{(1)} \frac{R_a^{(1)}}{2h_2} \ln\left(1 + \frac{2h_2}{R_a^{(1)}}\right)
\ee

\noi
where averaging has been performed from the apex ($\Delta = 0$) to a height $h_2$ above the apex.
Thus,

\be
\gamma_a^{(C)} \simeq  \gamma_a^{(2)} \langle \gamma_a^{(1)} \rangle \simeq  \gamma_a^{(2)} \gamma_a^{(1)} \frac{R_a^{(1)}}{2h_2} \ln\left(1 + \frac{2h_2}{R_a^{(1)}}\right).
\label{eq:main}
\ee

\noi
Eq.~(\ref{eq:main}) thus provides a useful approximation for the apex field enhancement factor of 2-primitive
compound structures when $h_2 < 0.2 h_1$. It can be generalized for $N$-primitive compound structures
and can be expressed as

\be
\gamma_a^{(C)} \simeq \gamma_N \Pi_{n=1}^{N-1} \gamma_a^{(n)} {\cal U}_n = \gamma_a^{CSC} \label{eq:CSC}
\ee

\noi
where 

\be
   {\cal U}_n = \frac{R_a^{(n)}}{2h_{n+1}} \ln\left(1 + \frac{2h_{n+1}}{R_a^{(n)}} \right)
\ee
   
\noi
We refer to Eq.~(\ref{eq:CSC}) as the Corrected Schottky Conjecture for small protrusions.
When $\frac{2h_{n+1}}{R_a^{(n)}} << 1$, a few terms in the expansion of the logarithm 

\be
 {\cal U}_n =  \left[ 1 - \sum_{k=1}^\infty  (-1)^{k-1} \frac{2^k}{k+1} \left(\frac{h_{n+1}}{R_a^{(n)}} \right)^k \right].
\ee

\noi
provides a useful approximation for ${\cal U}_n$:
Typically, for $h_{n+1}/R_a^{(n)} < 0.05$, less than 5 terms suffice.

\section{Numerical Verification}
\label{sec:numerical}

The Corrected Schottky Conjecture takes into account the  electrostatic field averaged over the
height of successive protrusions. It thus corrects the over-estimation of the apex field
enhancement factor of the compound structure. To see how effective this prescription is, we
consider the 2-primitive compound structures considered earlier.

\begin{figure}[hbt]
  \begin{center}
\vskip -0.75cm
\hspace*{-0.750cm}\includegraphics[width=0.575\textwidth]{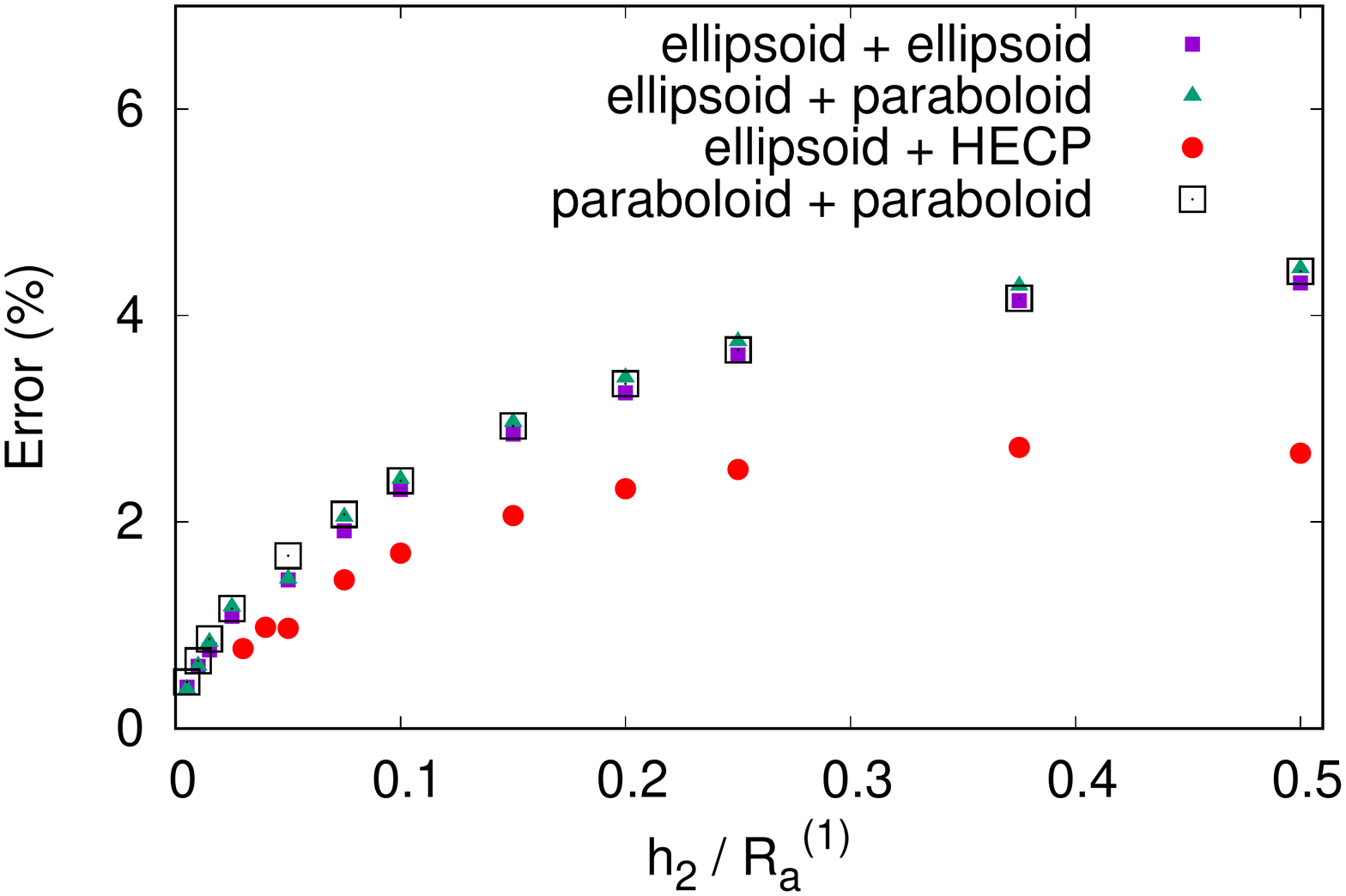}
\vskip -0.75cm
\caption{The relative error, as defined in Eq.~(\ref{eq:err2}),
  as a function of $h_2/R_a^{(1)}$ for the 4 compound shapes considered in Fig.~\ref{fig:error_schott}.
  The error in the Corrected Schottky Conjecture is small as compared to the
  Schottky Conjecture (see Fig.~\ref{fig:error_schott}).}
\vskip -0.5cm
\label{fig:error_csc_2prim}
\end{center}
\end{figure}

Fig.~\ref{fig:error_csc_2prim} shows the relative error in prediction using 
the Corrected Schottky Conjecture for the 2-primitive compound emitters used in Fig.~\ref{fig:error_schott}.
In this case (Fig.~\ref{fig:error_csc_2prim}), the relative error (\%) is defined as

\be
\text{Error} = \frac{\gamma_a^{(C)} - \gamma_a^{(CSC)}}{\gamma_a^C} \times 100  \label{eq:err2}
\ee

\noi
where the expression in Eq.~(\ref{eq:CSC}) is used for $\gamma_a^{(CSC)}$ while
$\gamma_a^{(C)}, \gamma_a^{(1)}$ and $\gamma_a^{(2)}$ have been obtained using COMSOL.
A comparison of
Figs.~\ref{fig:error_schott} and \ref{fig:error_csc_2prim} shows the vast
improvement in prediction of the Corrected Schottky Conjecture (CSC) especially at higher
values of $h_2/R_a^{(1)}$. Thus, Eq.~(\ref{eq:CSC}) can serve as a simple useful formula
to estimate the apex field enhancement factor of compound structures in terms
of the primitive components.

We next consider a 3-primitive structure (see schematic (f) of Fig.~\ref{fig:para_on_ellip})
to further ascertain the improved predictive capability
of CSC. The compound structure considered, consists of a hemiellipsoid as base with
$h_1 = 262.69\mu$m, $R_a^{(1)} = 200\mu$m, $\gamma_a^{(1)} = 3.33$, a paraboloid of height
$h_2 = 37.5\mu$m, $R_a^{(2)} = 1\mu$m, $\gamma_a^{(2)} = 23.33$
on top of the hemiellipsoid, and finally, a hemiellipsoid on cylindrical post (HECP) of height $h_3 = 100$nm,
apex radius of curvature $R_a^{(3)} = 10$nm and $\gamma_a^{(3)} = 10.05$ as the crown.

A direct application of the Schottky Conjecture predicts $\gamma_a^{(1)} \gamma_a^{(2)} \gamma_a^{(3)} \simeq 780.77$
as the apex enhancement factor of the compound structure while the actual value is $\gamma_a^{(C)} \simeq 572$.
The error is thus 36.5\%.
The Corrected Schottky Conjecture (Eq.~\ref{eq:CSC}) predicts 604.43 which is in error by about
5.67\%. Thus, the Corrected Schottky Conjecture for small protrusions 
predicts the apex enhancement factor of 3-primitive compound structures with
improved accuracy.

\section{Discussions and Conclusions}
\label{sec:discussions}

In the previous sections, we have proposed and verified a Corrected Schottky Conjecture (CSC)
which may be expressed as: {\it the apex field enhancement factor (AFEF) of a compound
structure consisting of $N$ primitive structures is approximately the
product of the crown AFEF and the product of the average AFEF of all the $N-1$ primitive-bases,
the average being over the height of the structure on top of each primitive-base}. If
the height of the structure on top of each primitive base is much smaller than the height of
the primitive base, the CSC may be expressed using a simple approximate formula given by Eq.~(\ref{eq:CSC}).

The Corrected Schottky Conjecture was found to be a useful approximation for determining the
apex field enhancement factor of compound structures under much relaxed conditions. The
effectiveness of Eq.~(\ref{eq:CSC}) for an $N$-primitive compound structure may be limited
by the approximate additive law of relative errors that is expected to hold
as indicated by the results of the 3-primitive example together with 
Figs. \ref{fig:error_schott} and \ref{fig:error_csc_2prim} 
for the original and corrected Schottky Conjecture. Thus, if $N$ is large,
the ratios $h_{i+1}/R_a^{(i)}$ must be small enough for the CSC of Eq.~(\ref{eq:CSC})
to have useful predictive capability. An aspect that has not been discussed so far involves
primitive structures with (finite sized) flat tops. Our numerical
studies show that the CSC in its generality (in terms of average enhancement factors) applies 
here as well within reasonable errors.

Finally, the Schottky Conjecture has sometimes been invoked to justify large values of
the field enhancement factor \cite{huang} as derived from experimental Fowler-Nordheim (FN) plots 
in field emission studies \cite{forbes15}. A couple of cautionary notes seem necessary. First, the
use of the elementary field emission equation \cite{forbes19,FN}
which uses the exact triangular barrier tunneling potential
results in unphysically high values of the enhancement factor to compensate for the
larger barrier height \cite{forbes19,db_rk19}. The actual enhancement factor required to explain an
experimental $I-V$ plot is always much lower when using a variant of the Murphy-Good \cite{MG,forbes2006,jensen_ency}
expression for the current density which uses the Schottky-Nordheim potential that
includes the image-charge contribution. For smooth curved emitter tips, the curvature
corrected expression for the current \cite{db_dist,db_rr19}
has been found to give results consistent with the physical measurements \cite{lee2018,db_rk19}
and may therefore be used directly to determine the value of the enhancement factor.

The second cautionary note follows from Fig.~\ref{fig:average_field}.
Note that $\gamma_a^{(C)}/\gamma_a^{(2)} \rightarrow 1$ as $h_2/R_a^{(1)}$ becomes large.
Thus a tall structure (such as a nanotube or nanowire) on top of a curved base
will not enjoy the multiplicative effect of the base as the field $E_z(z) \rightarrow E_0$
for $z$ large so that the average enhancement factor is close to unity. The Schottky Conjecture
thus cannot be used to justify the values of enhancement factors of compound
structure having a tall crown. This is of particular importance in multiscale
modelling of field emitters \cite{zhu2019}.

The Corrected Schottky Conjecture (CSC) provides a clearer picture about the domain
in which the multiplicative effect works and  it can help assess the 
role of micro-protrusions in experimental field emission results. The CSC can also be helpful
in designing experiments that aim to take advantage of multiplicative effects. For example,
a typical experimental cathode setup may consist of a metallic plate with a perpendicular
cylindrical stem on which single or multiple
wires may be mounted, each of which may have micro/nano protrusions or carbon nanowall/tube deposits. The
setup acts as a 3-primitive compound structure consisting of the stem, the wire and
the micro-protrusion. The height of the stem and the mounted wires may be adjusted
to maximize the multiplicative effect and increase the field emission current.

The analysis presented here deals with the apex field enhancement factor of a compound structure ($\gamma_a^{(C)}$)
consisting of protrusions from the apex of successive primitive bases. For such a geometric
arrangement, assuming the crown-apex to be smooth and locally parabolic, the field-emission
current can be easily determined using the generalized cosine variation of the electric
field \cite{db_ultram,cosine,db_rr19}. However, protrusions that grow away from the apex
need a separate analysis that seems challenging and is beyond the scope of the present manuscript.

\section{Acknowledgements}

The author acknowledges useful discussions with Gaurav Singh, Raghwendra Kumar and Rajasree Ramachandran
and anonymous referees for their comments that helped improve the manuscript.

\vskip -0.25 in
$\;$\\
\section{References} 



\begin{thebibliography}{99}
\bibitem{anode_position} When the anode is close to the apex, the local field
  at the apex depends on $h/D$. See \onlinecite{db_anode}. We shall
  consider the anode to be far away from the emitter tip in this paper.
\bibitem{see_miller} For certain special geometries amenable to conformal mapping,
such correction terms have been expressed in terms of the ratio of the height of the crown and the
width of the flat top of the base \cite{miller07,miller09}.
\end{thebibliography}
\end{document}